\documentclass{article}
\PassOptionsToPackage{table,dvipsnames}{xcolor}

\usepackage[utf8]{inputenc} 
\usepackage[T1]{fontenc}
\usepackage{hyperref} 
\usepackage{url}
\usepackage{booktabs}

\usepackage{amsfonts}
\usepackage{amsmath}
\usepackage{microtype}      
\usepackage{natbib}
\usepackage{tcolorbox}
\usepackage{amsthm}
\usepackage{subcaption}

\tcbuselibrary{skins, breakable}

\newtcolorbox{positionbox}{
  enhanced,
  breakable,
  colback=white,
  colframe=black!40,
  boxrule=0.6pt,
  arc=4pt,
  outer arc=4pt,
  attach boxed title to top left={xshift=12pt, yshift*=-0.5},
  title={\textbf{Position}},
  boxed title style={
    colback=white,
    colframe=black!40,
    boxrule=0.6pt,
    arc=3pt,
    outer arc=3pt,
    left=4pt, right=4pt,
    top=1pt, bottom=1pt,
    size=small,
  },
  left=6pt, right=6pt,
  top=10pt, bottom=6pt,
}
\usepackage[preprint]{neurips_2026}

\title{Healthcare LLM Benchmarks Are Only as Good as Their Explicit Assumptions}

\author{
Naveen Raman*\\
  Carnegie Mellon University\\
  \texttt{naveenr@cmu.edu}  \\
  \And
Santiago Cortes-Gomez*\\
  Carnegie Mellon University\\
  \texttt{scortesg@cs.cmu.edu} \\
\And 
Mateo Dulce Rubio*\\
  New York University\\
  \texttt{mateo.d@nyu.edu} \\
\And 
Fei Fang\\
Carnegie Mellon University\\
  \texttt{feifang@cmu.edu} \\
\And 
Bryan Wilder\\
Carnegie Mellon University\\
  \texttt{bwilder@cs.cmu.edu} \\
}

\begin{document}
\newcommand{\nrcomment}[1]{{\color{red} Naveen: {#1}}}
\newcommand{\mdr}[1]{{\color{blue} Mateo: {#1}}}

\maketitle

\begin{abstract}
Benchmarks are necessary for healthcare evaluation, but are not sufficient for predicting deployment performance. 
Our position is that the evaluation--deployment gap arises not because of poorly designed benchmarks, but from implicit assumptions about how users interact with models that cannot be surfaced from benchmarks alone. 
To make this precise, we propose a classification of assumptions into two categories: \textbf{task}, which can be tested from conversation data alone, and \textbf{outcome}, which requires outcome data and behavioral studies for testing. 
Critically, outcome assumptions depend on human behavior, something that even well-designed benchmarks cannot directly observe. 
To demonstrate the operationality of this framework, we retrospectively analyze a healthcare RCT as a case study and find that the gap naturally separates into task and outcome gaps of roughly equal size. 
To address this, we make two contributions: first, we propose \textit{BenchmarkCards}, an artifact that documents assumptions, and second, we propose \textit{staged evaluation}, a procedure that systematically tests assumptions and evaluates performance. 
\end{abstract}

\section{Introduction}
Healthcare LLM benchmarks are the dominant paradigm by which LLMs are evaluated prior to clinical settings, with high benchmark performance cited as preliminary evidence of clinical readiness~\citep{med_palm}. 
While benchmarks are an appropriate starting point for evaluation, LLMs are increasingly used to assist with patient health~\citep{self_diagnose_chat_gpt}, making it crucial that these models perform well in realistic settings. 
In contrast to sandboxable domains such as coding ~\citep{claude_code,cowork}, healthcare deployments necessarily contend with human interactions that are heavily context-dependent. 
As a result, benchmarks are a necessary indicator of model performance, but are not sufficient for predicting deployment performance. 
In fact, recent studies in healthcare have shown a large performance gap between evaluation and deployment in healthcare studies~\citep{nature_ai_medical_assistants,limitations_evaluation_medicine, nigeria_healthcare, medical_misinformation_chatbot}. 

Critically, our position is that \textbf{the evaluation--deployment gap arises not from poorly designed benchmarks but from implicit assumptions that separate evaluation from deployment.} 
Real-world complexities in clinical applications such as multi-turn interactions and noisy user queries are little reflected during evaluation. 
Moreover, some of the assumptions these benchmarks rely on, such as similarities between proxy and real outcomes, cannot be settled by any benchmark, no matter how well-designed. 
In other words, improving model performance on a benchmark will not lead to better deployment performance when the problem is about evaluation validity~\citep{hazard_rapid_approval,challenge_clinical_llm}. 
As a demonstration, we retrospectively apply this framework to a real-world clinical RCT, and find that modifications to benchmarks can only close around half the evaluation--deployment gap; closing the remainder requires real-world experiments such as RCTs. 

Prior work reframes evaluation as a validity or measurement problem~\citep{measurement_fairness,social_science_measurement}, but stops short of identifying which assumptions benchmarks can address, a key insight that dictates whether real-world experiments are necessary.
We add three ideas to operationalize this. 
First, we categorize assumptions into \textit{task}, which concern the relationship between conversations in the evaluation and deployment environments, such as differences in single vs. multi-turn interactions, and \textit{outcome}, which involve outcome data beyond the conversations themselves, such as the difference between proxy and clinical outcomes.
Critically, modifying benchmarks can only resolve task assumptions, and addressing outcome assumptions requires behavioral tests and RCTs because these assumptions rely on real-world outcomes.
Second, we propose \textit{BenchmarkCards}, a structured document which makes explicit the assumptions separating a benchmark from its intended deployment. 
\footnote{Code and schema for BenchmarkCards here: \url{https://github.com/naveenr414/benchmarkcards}}
Third, we propose a staged evaluation protocol for LLMs in healthcare, which starts from benchmark evaluation, then successively tests assumptions and re-evaluates performance.

\begin{figure}
\tikzstyle{mybox} = [draw=black!70, fill=Periwinkle!4, very thick,
    rectangle, rounded corners,
    inner sep=10pt]
\tikzstyle{fancytitle} = [rounded corners, fill=Periwinkle!20, text=black, very thick, draw=black!70]
\begin{tikzpicture}
\node [mybox] (box){%
    \begin{minipage}{0.9\columnwidth}
    \vspace{1.2mm}
Benchmarks in healthcare LLMs are necessary for evaluation but are insufficient to fully capture deployment performance. 
The core problem is not benchmark quality but implicit assumptions about task structure and decision-making. 
Leaving them unstated defeats the very purpose of evaluation: to estimate real-world performance. 
Closing the evaluation--deployment gap therefore requires making assumptions explicit, testing which assumptions hold through interaction data and behavioral studies, and updating evaluation protocols accordingly.
    \end{minipage}
};
\node[fancytitle, right=10pt] at (box.north west) {\textbf{Position}};
\end{tikzpicture}
\end{figure}
\section{Explicit Assumptions are Needed to Close the Evaluation--Deployment Gap}
\label{sec:current_procedures}

Evaluation in machine learning serves to quantify the performance of individual models and to compare performance across models~\citep{benchmarks_book}. 
While this paradigm has been instrumental in enabling the capabilities of modern systems, it is not self-evident that it suffices to assess deployment readiness. 
In high-stakes domains such as healthcare, evaluations must rigorously account for uncertainty and reproducibility to mitigate unintended side effects. 
Without such safeguards, shortcomings often surface only during or after deployment, with potentially severe consequences~\citep{are_we_learning}.
Inadequate evaluation has led to harmful outcomes in clinical trials, leading to the establishment of stringent evaluation standards that the AI community has yet to match~\citep{clinical_trials_complicated,hazard_rapid_approval,approval_pathway,thalidomide_scandal}.

Critically, in the healthcare LLM literature there is a growing body of evidence documenting a systematic gap between how LLMs perform in benchmark evaluation and how they perform in practice~\citep{nature_ai_medical_assistants, limitations_evaluation_medicine, nigeria_healthcare, medical_misinformation_chatbot}.
We argue that this gap is not merely a consequence of insufficient model capabilities or poorly designed benchmarks, but of assumptions that are implicitly embedded in evaluation protocols and inadvertently violated at deployment time. 
These assumptions concern how tasks are structured, who interacts with the model, and how its outputs translate into real-world decisions. 
When such assumptions are left implicit, strong benchmark performance can coexist with weak deployment performance, and practitioners have no systematic way to anticipate or mitigate this. 
Structurally, a gap between evaluation and deployment can arise due to various factors, including a distribution shift between real-world and controlled environments, differences in how users interact with the tool, and differences in objectives as benchmark evaluation typically optimizes for proxy outcomes. 
To make this concrete, we examine three representative studies from the healthcare literature, each of which illustrates a distinct way in which implicit assumptions drive the evaluation--deployment gap.
\begin{enumerate}
    \item \textbf{\citet{nature_ai_medical_assistants}} leverage LLMs to identify both diagnoses (e.g., conditions such as meningitis) and dispositions (e.g., decisions to either self-care, admit to hospital, etc.).  
    \textit{Benchmark performance:} Evaluated in standalone, single-turn prompts with complete patient information, LLMs achieve 95\% accuracy on diagnosing conditions and 56\% on selecting dispositions. 
    \textit{Deployment performance:} When used as assistive tools with human patients in back-and-forth conversation, performance drops to 34\% and 44\%, respectively. This is no better than situations where participants have no access to LLMs. 
    \textit{Assumption gap:} Benchmark evaluation assumed that real users present a single, clean prompt with full information and perfectly follow LLM instructions. 
    \item \textbf{\citet{limitations_evaluation_medicine}} compares the performance of LLMs on licensing medical exams against their performance on determining diagnosis with a real-world medical dataset.
    \textit{Benchmark performance:} LLMs perform well across a variety of licensing exams~\citep{what_disease_patient,llm_general_knowledge}, suggesting strong diagnostic ability.
    \textit{Deployment performance:} When facing patient scenarios from the MIMIC IV dataset~\citep{mimic_iv}, where LLMs are given information on patient symptoms, performance drops by up to 15\%. 
    \textit{Assumption gap:} 
    Licensing exams assume static interactions with all information provided upfront, while real clinical settings may involve incomplete and evolving information. 
    \item \textbf{\citet{nigeria_healthcare}} investigate the ability of LLMs to be effective assistants for healthworkers who test patients for conditions such as malaria. 
    \textit{Benchmark performance:} LLMs led health workers to rethink their test ordering and diagnoses for patients and also viewed the LLM-assisted notes as more favorable. \textit{Deployment performance:} LLMs did not improve healthworker ability to appropriately apportion tests to the necessary patients. 
    \textit{Assumption gap:} Evaluation relied on a proxy outcome, healthworker opinions, under the implicit assumption that favorable opinions translate into better diagnostic decisions.
\end{enumerate}

These three studies span different tasks, models, and healthcare settings, but each relied on an evaluation protocol encoding assumptions about task structure and decision-making that did not hold at deployment. 
Naturally, benchmarks are governed from different sets of assumptions that are an unavoidable part of any evaluation protocol---the problem is not that they exist, but that they are left implicit. 
When assumptions go unstated, the very purpose of benchmark evaluation, quantifying and comparing model performance to guide deployment decisions~\citep{benchmarks_book}, is defeated: practitioners have no way to assess whether benchmark results hold in their setting, or whether any available benchmark provides reliable guidance at all.

In healthcare, some assumptions, such as whether a proxy outcome reflects real-world impact, are simply not verifiable from benchmark data alone~\citep{high_stakes_decision_making, clin_consensus}. 
As a result, benchmarks are insufficient to close the gap, and instead we need to make assumptions explicit, so that practitioners can reason about when and where benchmark results transfer to deployment performance. 
In what follows, we propose a framework toward this goal.
\section{Understanding and Testing Assumptions}
\label{sec:instantiation}

\begin{figure}[t]
    \centering
    \includegraphics[width=\textwidth]{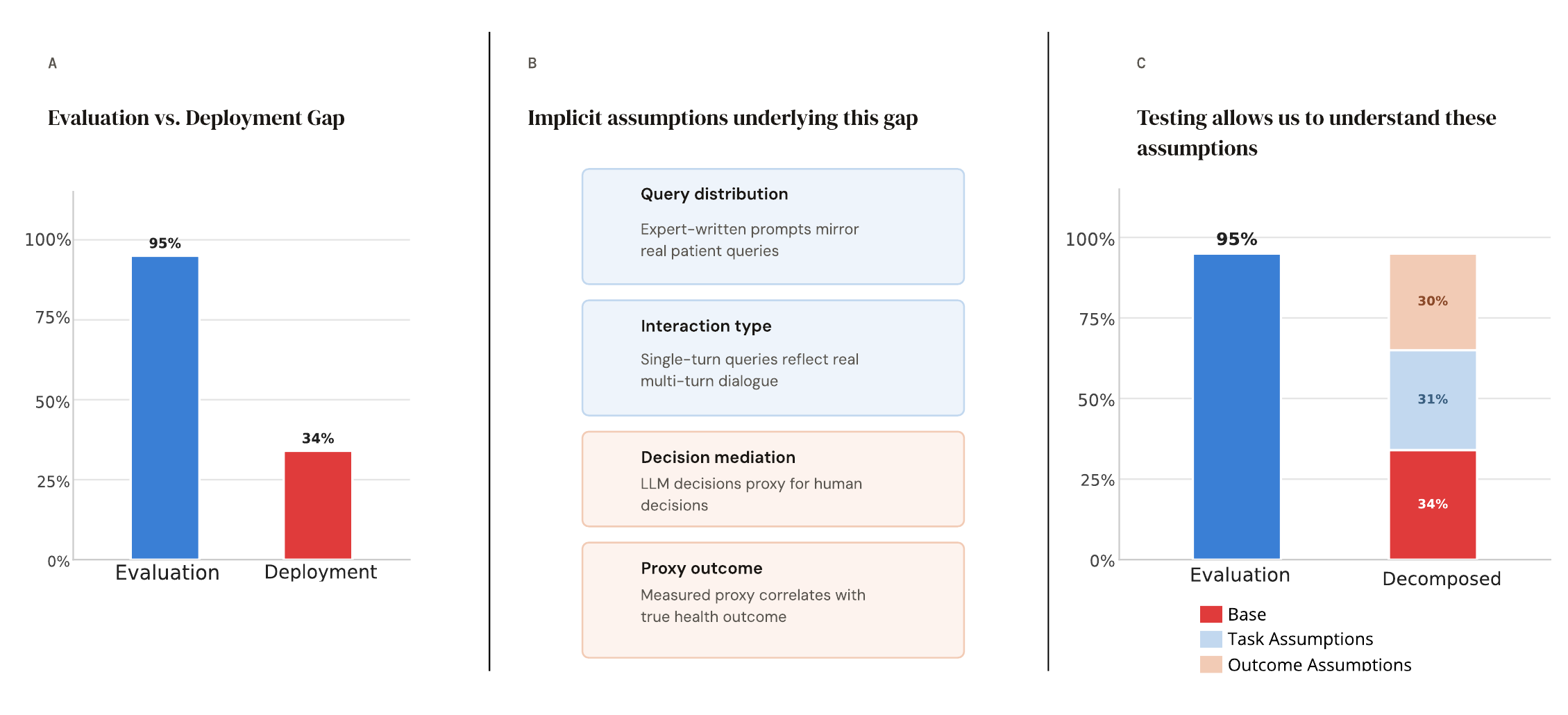}
    \caption{An illustration of how making assumptions explicit helps diagnose the evaluation--deployment gap. a) In~\citet{nature_ai_medical_assistants}, LLMs achieve 95\% performance during evaluation, but only 34\% during deployment.  b) This gap is driven by two task and two outcome assumptions. c) Sensitivity analysis allows us to decompose this gap and find that task assumptions are responsible for 31 percentage points and outcome assumptions for the remaining 30.}
    \label{fig:eval_card}
\end{figure}

\subsection{Formalizing the Assumptions Gap}
Let $R_{\mathcal{B}}(f)$ and $R_{\mathcal{D}}(f)$ be the benchmark and deployment performance for a model $f$ under some unspecified but fixed loss function. 
The evaluation--deployment gap is $R_{\mathcal{D}}(f)-R_{\mathcal{B}}(f)$.
As a diagnostic approximation, we can decompose the contribution of various assumptions $a$ as 
\[
R_{\mathcal{D}}(f)-R_{\mathcal{B}}(f) = \sum_{a} \Gamma_{a}.
\]
Such a decomposition is approximate because relaxing subsets of assumptions might produce different results, making the decomposition path-dependent. 
Our goal is not a precise computation of $\Gamma$ but a heuristic ranking of assumptions to understand which are most important to test. 
An assumption where $\Gamma_{a}$ is large for multiple paths indicates that it is robustly important, and so cannot be relaxed without a drop in performance. 
When all assumptions are satisfied, we have that $\Gamma_{a} \approx 0$ and so $R_{\mathcal{D}}(f) \approx R_{\mathcal{B}}(f)$. 
As we will explain in Section~\ref{sec:future}, the goal of staged evaluation is to confirm that all assumptions hold prior to deployment, rather than discovering this afterwards. 

\subsection{Testing Assumptions in Practice}
We propose a categorization of assumptions into two categories based on the data needed for testing: 
\begin{enumerate}
    \item \textbf{Task assumptions} are about whether the benchmark faithfully represents the conditions of deployment. Task assumptions revolve around conversation data, where well-designed benchmarks reduce dependence on task assumptions. 
    \item \textbf{Outcome assumptions} are about whether the benchmark's evaluation criterion tracks its decision-making target. Outcome assumptions depend on outcome data, such as what the patient does after interaction, and so cannot be tested or eliminated through benchmarks alone. 
\end{enumerate}

For both types, we note that passing a hypothesis test is necessary but not sufficient, and domain knowledge is still needed to assess whether the dimensions being compared are the correct ones.

Formally, let $P_{\mathcal{B}}$ and $P_{\mathcal{D}}$ denote the joint distributions over prompt-response pairs $(x, y)$ induced by the benchmark and deployment context respectively.
Then we define task assumptions to be those which can be tested through a hypothesis test $H_{0}: P_{\mathcal{B}} = P_{\mathcal{D}}$. 
Outcome assumptions involve data beyond what is captured in $P_{\mathcal{B}}$, hence why they cannot be improved through better benchmarks alone. 
Task and outcome assumptions informally map to internal and external validity, as task assumptions correspond to construct validity and concern whether benchmarks capture the correct phenomena, while outcome assumptions capture whether benchmarks correspond to real-world outcomes~\citep{quasi_experimentation}. 
Prior work~\citep{measurement_fairness,social_science_measurement, coston2023validity} identifies this measurement problem, and we build on this by describing the empirical procedure required by each type of assumption, making the framework actionable. 
   
\begin{table}[t]
  \centering
  \small
  \caption{Categorization of assumptions from prior work along with descriptions of their testability}
  \renewcommand{\arraystretch}{1.5}
  \label{tab:assumptions}
  \begin{tabular}{p{2.6cm}p{2cm}lp{3.5cm}p{2.5cm}}
  \toprule
  \textbf{Assumption Name} & \textbf{Papers} & \textbf{Type} & \textbf{How to Test} & \textbf{Data Required} \\ \midrule  
  Query distribution (doctor vs. patient-defined queries) & \citet{nature_ai_medical_assistants}, \citet{limitations_evaluation_medicine} & Task & Collect queries from patients and doctors and compare similarity in distribution. & Sample Deployment Queries \\
  Interaction type (single- vs. multi-turn) & \citet{nature_ai_medical_assistants}, \citet{limitations_evaluation_medicine} & Task & Collect data under static and dynamic interaction patterns and compare performance. & Multi-turn interactions  \\
  Decision Mediation (user vs. LLM makes the decision) & \citet{nature_ai_medical_assistants} & Outcome & Compare decisions driven by LLMs against decisions made by users. & Behavioral experiment with decisions made by users \\
  Proxy Outcome (proxy vs. clinical) & \citet{nigeria_healthcare} & Outcome & Compute the relationship between a proxy and the true outcome & Either proxy vs. clinical outcome relationship in the literature or a new RCT. \\ 
  \bottomrule
  \end{tabular}
  \end{table}

Table~\ref{tab:assumptions} illustrates that task assumptions typically deal with similarities in queries in different scenarios, while outcome assumptions deal with decision-making impacts beyond the queries. 
In practice, this means that task assumptions are about distributions that can be sampled and analyzed, which contrasts with outcome assumptions that deal with outcomes. 
Critically, data collected to test either class of assumptions is diagnostic rather than evaluative, and not at benchmark-scale. 
We next discuss how we can understand \textit{which assumptions} are most important for a given deployment context.

\subsection{Understanding Which Assumptions are Most Important}
\label{sec:important_assumptions}
The practical value of making assumptions explicit depends on understanding their relative importance for a given deployment context. 
For instance, relaxing some assumptions can yield a large gap between benchmark and deployment performance, while others can have negligible impact.
Several tools exist for reasoning about this, including red teaming~\citep{red_teaming_silver_bullet}, partial identification \citep{cortes2024statistical}, and sensitivity analysis from the causal inference literature \citep{luedtke2015statistics}. 
We focus on sensitivity analysis here, and show an example of how it could be used to estimate values of $\Gamma$. 
We focus on~\citet{nature_ai_medical_assistants} as an illustration because it is one of the few studies to contain data on a large-scale RCT with real patient interactions (see Figure~\ref{fig:eval_card}). 

In detail, consider two task assumptions embedded in the benchmark from~\citet{nature_ai_medical_assistants}: query distribution (doctor-written vs. patient-written queries) and interaction type (single-turn vs. multi-turn).
Starting from the original benchmark performance of 95\%, we relax each assumption in turn.
We note that while the path matters for the individual contributions of each assumption, \textit{the overall contribution of task assumptions remains constant independent of ordering}. 
Replacing doctor-written queries with patient-written single-turn queries reduces performance to 83\%, attributing $\Gamma = 12$ percentage points to query distribution. Next, we relax interaction type by comparing performance from single-turn and multi-turn user queries, and find that performance drops from $83\%$ to $64\%$, attributing an additional $\Gamma = 19$ percentage points to interaction type. Together, task assumptions account for approximately 31 percentage points of the total gap.
In the RCT with real users, the performance drops from $95\%$ to $34\%$, so that the remaining 30 percentage points difference can be attributed to outcome assumptions. 
This demonstrates how decompositions can be empirically analyzed, though the numbers from this case study do not necessarily generalize.
We next show how insights into testing assumptions can guide how we conduct evaluations.
\footnote{Code for analysis here: \url{https://github.com/naveenr414/healthcare-llm-eval-gap}}

\section{Implications for Evaluation}
\label{sec:future}
\subsection{Documenting Assumptions through \textit{BenchmarkCards}}

A first step towards making assumptions explicit is standardizing how assumptions are documented. 
With this aim, we propose \emph{BenchmarkCards} as structured documentation that designers report together with their benchmarks, explicitly stating the assumptions about deployment contexts encoded in their evaluation protocol, including how tasks are structured and how decisions are made.
BenchmarkCards build on existing documentation practices widely adopted in the AI community including Model Cards \citep{model_cards}, EvalCards \citep{eval_cards}, and Datasheets for Datasets \citep{gebru2021datasheets}. 
Unlike Model Cards and EvalCards, which are model-centric, capturing a model's performance and properties, BenchmarkCards are benchmark-centric and agnostic to specific modeling and deployment decisions. Rather than describing a particular model, they describe what a benchmark does and does not capture, making transparent the assumptions about deployment contexts and the extent to which benchmark evaluations translate into real-world performance.

The card is structured around the dichotomy between task and outcome assumptions, documenting how tasks are structured and how decisions are made in the intended deployment context. 
Benchmark designers fill out BenchmarkCards by answering questions about their evaluation protocol, without needing to anticipate any particular downstream use. 
Table~\ref{tab:benchmark_card} (left) shows examples of such questions, each with a factual answer grounded in design choices rather than any specific deployment instance. 
An additional example of a BenchmarkCard is provided in Appendix~\ref{sec:second_benchmark_card}.

A practitioner facing a deployment decision then uses BenchmarkCards to assess which assumptions hold in their setting (Table~\ref{tab:benchmark_card}, right), and identifies which benchmarks most closely reflect their use case. The closest-matching benchmarks can then guide model selection, pointing to the model that performs best under the most similar evaluation conditions. Critically, if no existing benchmark closely matches the deployment context, BenchmarkCards make that gap visible rather than hiding it, supporting the harder question of whether any available benchmark provides reliable guidance at all.

\begin{table*}[t]
  \centering
  \small
  \caption{BenchmarkCard (left, filled once by benchmark designers) and practitioner deployment assessment (right, filled per deployment context) for~\citet{nature_ai_medical_assistants}.}
  \label{tab:benchmark_card}
  \renewcommand{\arraystretch}{1.5}
  \begin{tabular}{p{2.6cm}lp{4.2cm}@{\hspace{0.4cm}}!{\vrule}@{\hspace{0.4cm}}p{3.1cm}}
  \toprule
  \textbf{Question} & \textbf{Assumption} & \textbf{Answer} & \textbf{Holds at deployment?} \\
  \midrule
  What is the intended use case? & -- & LLM performance in identifying diagnosis and dispositions. & Use case: patients interact with LLM as medical assistants. \\
  Who created the examples, and what are their backgrounds? & Task & Seven physicians with backgrounds in general practice and intensive care, who averaged 24 years of experience. Three created questions; four reviewed. & \textbf{No.} Patients generate queries themselves; expert curation does not reflect this. \\
  Are examples information--complete? & Task & Yes. All examples contain sufficient information to answer the query. & \textbf{Partially.} Patients may omit or misreport information. \\
  Are examples single or multi-turn? & Task & Single-shot prompts only. & \textbf{No.} Deployment features multi-turn interactions. \\
  Does the benchmark treat LLM output as the final decision? & Outcome & Yes. The LLM decides directly; no human mediator is present. & \textbf{No.} Users make the final decisions, not the LLM. \\
  What outcome is measured? & Outcome & Rate at which the LLM selects the correct diagnosis or treatment. & \textbf{No.} Real outcome depends on user uptake, not LLM accuracy alone. \\
  \bottomrule
  \end{tabular}
\end{table*}

BenchmarkCards can also signal to the community where additional efforts are needed to construct new benchmarks. Recurring gaps in query distribution or interaction type point to the evaluation conditions that are most underrepresented, such as the multi-turn gap that new benchmarks in the healthcare domain are already investigating~\citep{multi_turn_survey}. 
In this way, BenchmarkCards advance safe and transparent LLM deployment, and open new research avenues relevant for real-world use.

Incorporating BenchmarkCards into the practices surrounding conferences would help improve the rate of adoption. 
For example, benchmark and dataset tracks at conferences could require completion of BenchmarkCards during submission, along with a discussion of potential implicit assumptions. 
BenchmarkCards could also become a common artifact shipped with benchmarks on sites like HuggingFace, thereby making it easier for practitioners to identify which benchmarks differ from their existing practice. 
To ensure ease of integration, such ideas could also be incorporated into the datasheets that accompany datasets~\citep{gebru2021datasheets}.
Each of these align practitioner and benchmark designer incentives so that transparency in benchmarks becomes standard.

\subsection{Staged Evaluation Protocols}
We outline a ``staged'' evaluation protocol where practitioners iteratively test assumptions and use this to guide evaluation. 
We provide an overview of what evaluation would look like below, then discuss how this applies to~\citet{nature_ai_medical_assistants}. 
\begin{enumerate}
    \item \textbf{Compare BenchmarkCards against Deployment} - After evaluating performance on some selected benchmark, read through the BenchmarkCard to understand what the gaps are between evaluation and deployment.
    \item \textbf{Collect Data for Task Assumptions} - Identify which task assumptions can be tested and collect the appropriate data needed for the test (see Table~\ref{tab:assumptions}). For example, collect data on real user interactions to capture the difference in query distribution. 
    \item \textbf{Test Task Assumptions} - After data collection, test task assumptions by understanding performance degradations between evaluation and deployment. 
    For assumptions that have large performance drops, improve model performance potentially through the collection of more \textit{targeted} data. 
    Once task assumptions are either all verified or have sufficiently small performance gaps (as determined by practitioners), proceed to outcome assumptions. 
    \item \textbf{Rank Outcome Assumptions} - Using domain expertise, understand which outcome assumptions are most important and which need to be feasibly evaluated because evidence is not present in the literature. 
    For assumptions that lack evidence from the literature and can have a large impact on performance, testing is necessary. 
    \item \textbf{Test Outcome Assumptions} - Run behavioral studies or RCTs for the most important outcome assumptions. 
    For assumptions that fail, identify whether modifications to the model or the evaluation procedure can help bridge the gap. 
    For example, if proxy outcomes differ from clinical outcomes, then practitioners can either measure the clinical outcome directly, or find an alternative valid proxy, which is well-documented in the clinical trials literature~\citep{fda_bio_markers_endpoints}. 
    After the performance drop due to unsatisfied assumptions is sufficiently low, evaluation and deployment performance should be approximately equal, and practitioners can safely deploy. 
\end{enumerate}

Returning to the example from~\citet{nature_ai_medical_assistants}, there are both task assumptions, which are responsible for a 31 percentage points reduction in performance, and outcome assumptions, which are responsible for the remaining 30 percentage points drop. 
The former can be addressed through modification of the benchmark to cover both patient-written queries and multi-turn interactions.
Addressing the latter requires behavioral studies to identify the rate at which patients listen to LLM-recommended dispositions. 
If patients follow LLM-recommended dispositions at high rates, then measuring the quality of LLM-recommended dispositions suffices to evaluate deployment performance; if not, then we would need to either find a valid proxy or explicitly evaluate with patient decisions. 
In some situations, practitioners can sidestep proxies by modifying the setup so the assumption holds true (e.g., having some reminder system in place so user-mediated decisions match LLM-mediated decisions). 

Finally, a database of LLM trials and assumptions allows practitioners to identify which assumptions have previously been tested, which could reduce the burden of running staged evaluations. 
In clinical trials, many studies and evaluations are preregistered to an appropriate database, and can be queried for different properties~\citep{clinical_trials_gov}. 
Similar efforts must be made for LLM trials, particularly 1) a database with different LLM RCTs across domains, and 2) a database of different behavioral and real-world evaluations to assess which assumptions hold in which domains. 
The former can help with the design and analysis of evaluation protocols, and a searchable database with this information would assist practitioners looking to assess the impact of LLMs in similar domains. 
The latter assists practitioners in estimating the importance of assumptions. 
Together, BenchmarkCards, staged evaluation, and an assumptions database create a pathway towards evaluation that centers around assumptions and closes the gap between evaluation and deployment. 
\section{Related Work}
\paragraph{Evaluation Validity.} 
A large body of literature studies the validity of evaluation by characterizing different types of validity issues that can arise~\citep{research_reliability,quasi_experimentation}. 
Within machine learning, prior work has tackled issues of whether current evaluation platforms are ecologically valid, including in healthcare~\citep{medical_llm_construct_validity}, benchmarks~\citep{real_world_validity,everything_in_the_world_benchmark}, and generative AI~\citep{validity_gen_ai_eval}.
Such a line grew out of a line of work criticizing the narrow scope of machine learning evaluations~\citep{evaluation_gaps_machine_learning_practice, coston2023validity}. 
Most related is a line of work reframing machine learning evaluations under a social science perspective, including a discussion of the role that assumptions play~\citep{measurement_fairness,social_science_measurement}. 
While they establish benchmark validity as a measurement problem, they do not distinguish between assumptions that require conversation data vs. those that require real-world outcome data. 
We address this issue by separating assumptions into task and outcome, which is essential because it determines how much of the gap can be closed through better benchmarks.

\paragraph{Criticisms of Benchmarks.} 
A variety of prior work has critiqued the standard benchmark-driven approach broadly in machine learning. 
Critiques include those who argue that Goodhart's law leads us to over-optimize for benchmarks at the expense of real capabilities~\citep{goodhart_law_ml}, whether benchmarks can ever really be truly ``representative'', and whether that is broadly something to even strive for~\citep{everything_in_the_world_benchmark}. 
While these works focus on benchmark quality, our position is that even well-designed benchmarks cannot tackle certain gaps between evaluation and deployment, requiring changes to the way we think about assumptions. 

\paragraph{Clinical Trials.}
The literature on clinical trials studies how clinical procedures or new drugs can be verified. 
The gold standard is running RCTs to determine effectiveness, and in the United States, the FDA sets strict guidelines on how clinical trials should be run and their different phases~\citep{clinical_trial_phases}. 
While involved, clinical trials require such complications because they need to establish the safety of new drugs beyond doubt; failure to run thorough clinical trials can result in large safety risks~\citep{clinical_trials_complicated,thalidomide_scandal}. 
Because establishing improvement on clinical outcome measures can be expensive or even infeasible, some trials use surrogate outcomes instead as a measure of success~\citep{fda_bio_markers_endpoints}. 
Surrogates allow for cheaper evaluation but run the risk of an inappropriate endpoint built on correlative evidence. 
As a result, a large literature has developed around \textit{how} surrogates should be selected and evaluated~\citep{surrogate_criteria}.
Our work translates principles from clinical evaluation into LLMs in healthcare, with the insights on surrogate outcomes and trial design influencing how best to structure evaluation, and the staged evaluation mimicking the phases of clinical trial evaluation. 
\section{Alternate Viewpoints}
\textbf{Alternate Viewpoint 1:} Testing assumptions and conducting sensitivity analyses is too costly of a procedure to be done practically. 

\textit{Rebuttal:} Testing task assumptions via sensitivity analyses requires only a few samples from the deployment context, while non-testable assumptions can be reasoned based on domain knowledge. 
As a result, the cost is bounded and front-loaded. 
Verified assumptions reduce evaluation costs because if assumptions hold, simpler protocols suffice, while non-verified assumptions pinpoint the exact protocols necessary for evaluation. 

\textbf{Alternate Viewpoint 2:} Benchmarks have been responsible for much of the success in machine learning in the past few decades; why should healthcare be any different? 

\textit{Rebuttal:} In sandboxable domains such as coding, evaluations are representative of deployment.
However, evaluation cannot be sandboxed in healthcare because it relies on patient behavior which is difficult, if not impossible, to capture entirely through benchmarks. 
Instead, benchmarks represent a static snapshot at one point in time, while physician and patient interactions are heterogeneous in time and location. 
Frameworks that make assumptions explicit allow us to dynamically analyze this drift between evaluation and deployment in a principled manner that static benchmarks could not capture. 

\textbf{Alternate Viewpoint 3:} Making assumptions explicit only codifies knowledge that is already obvious to practitioners.

\textit{Rebuttal:} Due to the advanced capabilities of modern LLMs, practitioners from diverse backgrounds often use the same model for fundamentally different tasks. These differences imply distinct deployment contexts. Given this diversity, it is not reasonable to assume that practitioners share the same preconceptions that will lead towards the same set of implicit assumptions required to interpret benchmarks and bridge the training–deployment gap.
Further, sensitivity analysis formally quantifies which assumptions are most important, something which practitioners cannot determine from implicit knowledge. 
For example, while practitioners might know that patient queries are different from those in a benchmark, what is not clear is whether that assumption is more important than the difference between single-turn and multi-turn interactions. 
Understanding the relative importance of assumptions allows for the design of evaluation protocols around the most important assumptions to ensure that evaluations are representative of deployment. 

\textbf{Alternate Viewpoint 4:} Evaluations that make assumptions explicit create additional burden with little regulatory benefit. 

\textit{Rebuttal:} Frameworks such as BenchmarkCards delineate when benchmarks should and should not be used, and the circumstances under which benchmarks remain valid. 
This supports regulatory bodies as it allows for an honest assessment of where liability should be.
Moreover, staged evaluation can naturally be related to the clinical trials literature, where evaluation occurs with sequential trials in increasing order of complexity and size. 
There are also analogs to the use of proxies and simplifying assumptions in that literature, as surrogate endpoints can be used in situations where they are validated and assessing the true outcome would take too long or be too costly~\citep{fda_bio_markers_endpoints}.
Regulatory bodies have shown flexibility in balancing efficiency and validity when pursuing a maximally efficient trial design is too inefficient. 
Similarly, staged evaluation does not require that all assumptions are fully addressed, but rather that non-verified assumptions do not significantly create an evaluation--deployment gap. 

\textbf{Alternate Viewpoint 5:} LLM capabilities will eventually match healthcare deployments, rendering frameworks that make assumptions explicit unnecessary. 

\textit{Rebuttal:} Regardless of model capability, proper evaluation is necessary in high-stakes domains such as healthcare. 
Current evaluation procedures are too far removed from reality to guarantee safe deployment; performance on benchmarks does not equate with performance in reality. 
Instead, assumptions-explicit frameworks break down the gaps between evaluation and deployment so models can iteratively be tested closer to real clinical settings. 
\section{Limitations and Conclusion}
\paragraph{Limitations.} Our sensitivity analysis is done with one particular paper; the purpose of this is not as a definitive characterization of the importance of different assumptions, but rather an illustration of what sensitivity analysis might look like. 
BenchmarkCards are proposed but not validated; it remains an open question to see what impact BenchmarkCards will have on deployment. 
Finally, our work presents task vs. outcome as a binary decision when in reality different assumptions vary continuously along a spectrum. 
For example, some assumptions might require both outcome and conversational data, and other outcomes might not be testable even with access to outcome data. 
Moreover, some outcomes might only be testable over long timescales, making it necessary to find reasonable proxies. 

\paragraph{Extensions Beyond Healthcare.} Our position focuses on healthcare by design, but many of the same ideas are applicable to any field where deployment is costly and detached from evaluation. 
Our assumptions are largely motivated by healthcare, but it would be of interest to see what types of modifications are needed for extensions to fields such as finance~\citep{llm_finance} and law~\citep{llm_legal}. 
Such fields are similar in that decisions cannot be sandboxed and necessarily interact with real people or markets. 
Therefore, similar frameworks of assumption testing and staged evaluation are necessary to understand performance in a methodical manner. 

\paragraph{Conclusion.} Better benchmarks are necessary but insufficient for deploying LLMs in healthcare. 
Our position is that closing the gap between evaluation and deployment requires making explicit the assumptions separating the two. 
We propose classifying assumptions into two categories: task and outcome, which differ based on whether they can be tested purely from conversations. 
A case study on a real-world RCT reveals that outcome assumptions account for around half the gap between evaluation and deployment, implying that well-designed benchmarks cannot fully close the gap. 
Instead, real-world behavioral tests and RCTs are needed to understand whether outcome assumptions hold. 
To improve evaluations, we first propose BenchmarkCards for better documenting assumptions, and we second propose staged evaluation as a procedure to iteratively test assumptions and update evaluations accordingly. 
Evaluation in healthcare suffers from a transparency problem that is solvable by making explicit the assumptions that separate evaluation and deployment.

\newpage 
\bibliographystyle{plainnat}
\bibliography{references}

\newpage 

\appendix 
\section*{Acknowledgements}
We thank Lawrence Jang, Amanda Coston, Luke Guerdan, and Sang Truong for their comments on this work. 
This work was supported by the National Science Foundation (NSF) program on Civic Innovation Challenge (CIVIC) under Award No. 2527408. 
Co-author Raman is supported in part by an NSF GRFP award. 

\section{A Second BenchmarkCard}
\label{sec:second_benchmark_card}
We fill out a BenchmarkCard for~\citep{limitations_evaluation_medicine} to demonstrate the generality of BenchmarkCards. 
The BenchmarkCard can be found in Table~\ref{tab:benchmark_card_hager}.

\begin{table*}[t]
  \centering
  \small
  \caption{BenchmarkCard (left, filled once by benchmark designers) and practitioner deployment assessment (right, filled per deployment context) for~\citet{limitations_evaluation_medicine}, where the benchmark is licensing exams and deployment is clinicians from MIMIC IV~\citep{mimic_iv}.}
  \label{tab:benchmark_card_hager}
  \renewcommand{\arraystretch}{1.5}
  \begin{tabular}{p{2.6cm}lp{4.2cm}@{\hspace{0.4cm}}!{\vrule}@{\hspace{0.4cm}}p{3.1cm}}
  \toprule
  \textbf{Question} & \textbf{Assumption} & \textbf{Answer} & \textbf{Holds at deployment?} \\
  \midrule
  What is the intended use case? & -- & LLM guidance to clinicians on which tests to run, based on licensing exam questions. & Use case: clinicians interact with LLMs iteratively to guide diagnosis and test ordering. \\
  Who created the examples, and what are their backgrounds? & Task & Examples come from central licensing boards, generally consisting of clinicians and teachers. Specific dates are not mentioned. & \textbf{Partially.} Licensing boards have clinical expertise, but queries differ from real patient scenarios. \\
  Are examples information-complete? & Task & Yes. All examples contain sufficient information to be answered based only on the query. & \textbf{No.} In practice, clinicians do not have access to all information upfront and must interact iteratively. \\
  Are examples single or multi-turn? & Task & Single-shot prompts only. & \textbf{No.} Deployment features multi-turn interactions between clinicians and LLMs. \\
  Does the benchmark treat LLM output as the final decision? & Outcome & Yes. The LLM makes the final decision directly. & \textbf{No.} Physicians retain discretion over which tests to run and how to diagnose patients. \\
  What outcome is measured? & Outcome & Accuracy on licensing exam questions. & \textbf{No.} Real outcome is diagnosis ability and lab test interpretation on actual patient cases. \\
  \bottomrule
  \end{tabular}
\end{table*}


\end{document}